\documentclass[pra,twocolumn,superscriptaddress,nobibnotes]{revtex4-2}
\usepackage[utf8]{inputenc}
\usepackage{amsmath,bm}
\usepackage{amsfonts}
\usepackage{amssymb}
\usepackage{xcolor}
\usepackage{braket}
\usepackage{graphicx}
\usepackage{subfigure}
\usepackage{todonotes}
\usepackage{subfigure}
\usepackage[colorlinks=true, allcolors=blue]{hyperref}
\usepackage{dsfont}

\begin{document}

\title{Voltage-tunable, femtometer-precision plasmo-mechanical displacement \\ at fixed gap size}

\author{Rasim Volga Ovali}
\affiliation{Recep Tayyip Erdogan University, 53100 Rize, Turkey}
\author{Mehmet Emre Tasgin}
\affiliation{Institute of Nuclear Sciences, Hacettepe University, 06800 Ankara, Turkey}

\date{\today}

\begin{abstract}
	We propose an elegant method for continuous electrical-tuning of plasmo-mechanical displacement and squeezing without changing plasmonic gap size. Recent experiments bend the mechanical oscillator~(cantilever) in units of nm via electrostatic actuators. We do not bend the cantilever but merely electrically-tune the gap intensity, so plasmo-mechanical coupling, via Fano resonance. This allows continuous displacement tuning in units of mechanical oscillator length that is about 30 fm in the experiments. This way, coupling strength can be tuned by 2 orders-of-magnitude via only a 1 V potential difference. Response time is \textit{picoseconds}. Moreover, quadrature-squeezing~(entanglement) of the oscillator can also be tuned continuously.   
\end{abstract}

\maketitle

\section{Introduction}

Coupling of a mechanical oscillator with light~(optomechanics) possesses important implementations such as optical cooling of the mechanical motion~\cite{bhattacharya2007trapping}, transfer of optical quantum states into mechanical ones~\cite{zhang2003quantum}, and for introducing light-controlled quantum superposition states~\cite{meystre2013short} in the mechanical motion which made, for instance, detection of gravitational waves~(LIGO) possible~\cite{jia2024squeezing}.

Plasmo-mechanics, a newly developing field, has the potential to carry these phenomena into smaller dimensions~\cite{thijssen2013plasmon,roxworthy2016nanomechanical,ou2016giant,zhu2016plasmonic,dong2016gigahertz,roxworthy2018electrically,haffner2019nano,xu2021recent,koya2021plasmomechanical}. This field provides both controlablity and integrability that are necessary for implementations with quantum circuits. Localized surface plasmons~(LSPs), appearing at nm-sized metallic gaps, take the place of photons. LSPs confine light into hotspots~(gap) where intensity can be enhanced by several orders of magnitude~\cite{roxworthy2016nanomechanical,roxworthy2018electrically,haffner2019nano}. In plasmomechanics, the near-field excitations (plasmons) create the radiation pressure in analogy to photons. Plasmon--mechanical interaction becomes greatly enhanced compared to the conventional optomechanics.

Plasmonics can also provide control over the plasmo/opto-mechanical interactions. For instance, recent state-of-the-art experimental works demonstrate the electrical control of the plasmonic gap size, so the LSP resonances and the gap intensity~\cite{roxworthy2018electrically,haffner2019nano}. The LSP resonance appears in the gap between a cantilever, into which a cuboid gold nanostructure is embedded, and a gold pad lying at the bottom. See Fig.~1. Voltage applied on electrostatic actuator bends the equilibrium position of the cantilever, so the gap size changes between 20 nm--70 nm. This way, effective plasmomechanical coupling can be tuned substantially within a time of 10 nanoseconds. However, bending of the cantilever  avoids the electrical-tuning of the displacement in the order of cantilever's oscillator strength, $x_{\rm \scriptscriptstyle ZPF} \approx$30 fm in Ref.~\cite{roxworthy2018electrically}, but allows a nm-size tuning only.

In this paper, we show that electrical tuning of gap plasmon intensity~(so the effective plasmomechanical coupling~\cite{genes2008robust,vitali2007optomechanical}) can be achieved without bending the cantilever. The cantilever displacement is solely controlled by the radiation pressure of the gap intensity. This makes tuning of the displacement in units of $x_{\rm \scriptscriptstyle ZPF} \approx$30 fm possible. Response time is \textit{picodeconds}. We utilize a Fano resonance~(transparency)~\cite{leng2018strong,wu2010quantum,shah2013ultrafast,tacsgin2018fano,limonov2017fano} induced by a layer of defect-centers (relies in the LSP gap) to turn off the gap intensity. See Fig.~1. We demonstrate the phenomenon via FDTD simulations using Lumerical. Application of only a 0.8 V potential difference is sufficient to continuously tune the gap intensity between 1--200. See Fig.~4.
Our scheme is also compatible with sideband cooling~\cite{bhattacharya2007trapping}~\cite{PS_sideband}. 
Not only the displacement but also quadrature-squeezing~(can also be converted into entanglement~\cite{ge2015conservation,von2024engineering}) of the cantilever oscillations is tuned continuously. See Fig.~5. 
 We use the experimental parameters~\cite{roxworthy2018electrically} in our  calculations. 

The plasmomechanical system (Fig.~1) is excited by a laser of frequency $\omega$~($\lambda=c/\omega$). A transparency takes place at the gap intensity~(LSP) due to the Fano resonance~(FR)~\cite{leng2018strong,wu2010quantum,shah2013ultrafast,tacsgin2018fano,limonov2017fano}. Please compare Figs.~\ref{fig2} and \ref{fig3}. FR is induced by a densely-implanted defect-centers~\cite{murata2011high,rotem2007enhanced,ngandeu2024hot} layer~(quantum emitter, QE). Such a transparency, which avoids also the optomechanical coupling, takes place when $\omega$ is around the level-spacing of the defect-centers $\Omega_{\rm \scriptscriptstyle QE}$, i.e., for $\omega=\omega_{\rm f} \sim \Omega_{\rm \scriptscriptstyle QE}$~\cite{PSwf}. The voltage applied on the defect-centers shifts the level-spacing $\Omega_{\rm \scriptscriptstyle QE}$ with respect to the excitation frequency $\omega$. Thus, the number of plasmons in the gap~(Fig.~4), so the optomechanical coupling and the displacement~(Fig.~5a), are continuously tuned.

\begin{figure}
	\centering
	\includegraphics[width=1\linewidth]{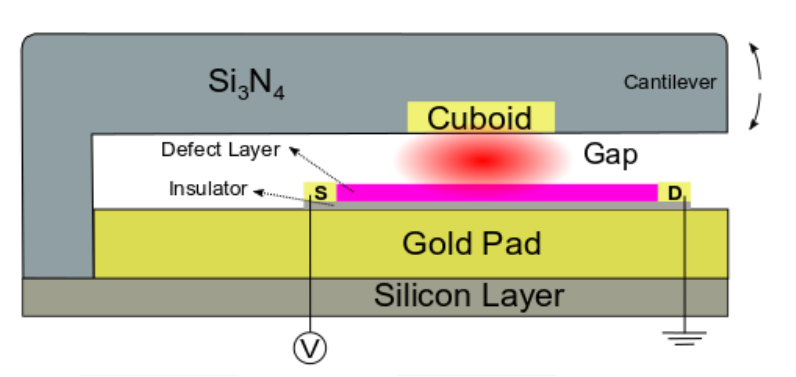}
	\newline
	\caption{Fano-control of plasmomechanical system. A gold cuboid of dimensions 20$\times$20$\times$40 nm is embedded into the bottom side of a ${\rm Si}_3 {\rm N}_4$ cantilever vibrating at frequency $\omega_m=$10 MHz~\cite{roxworthy2016nanomechanical,roxworthy2018electrically}. A gap plasmon resonance occurs between the cuboid and the gold pad lying 20 nm below the cantilever. Incident light of frequency $\omega$ is along the z-direction. A layer of densely-implanted defect-centers~\cite{murata2011high,rotem2007enhanced,ngandeu2024hot} of resonance $\Omega_{\rm \scriptscriptstyle QE}$ is placed on the pad. It creates a Fano transparency in the gap intensity at about the defect-resonance, i.e.,  $\omega=\omega_{\rm f}\sim \Omega_{\rm \scriptscriptstyle QE}$ ~\cite{PSwf}. The potential difference applied on the layer electrically-tunes the defect resonance~\cite{larocque2024tunable,shibata2013large,miller1985electric}, so also continuously tunes the gap intensity. The cantilever is not bended in difference to Refs.~\cite{roxworthy2018electrically,haffner2019nano}.
	   }
	\label{fig1}
\end{figure}

FR is the plasmon analog of electromagnetically-induced transparency~(EIT) formerly studied on atomic vapors~\cite{fleischhauer2005electromagnetically}. Metal nanostructures~(MNSs) localize the incident light on nm-sized hotspots~(gaps) as plasmonic oscillations. When a QE is placed at one of the hotspots, a strong plasmon-QE coupling takes place. This introduces a narrow transparency window in the plasmon spectrum~\cite{leng2018strong,wu2010quantum,shah2013ultrafast,tacsgin2018fano,tacsgin2013metal}. A FR takes place in a narrow frequency band which makes it unuseful for broadband applications~\cite{PSDarkmodes}. Such a disadvantage, however, turns out to be highly beneficial for electrical tuning of linear/nonlinear optical processes~\cite{gunay2020continuously,polat2024continuous} which is not available by other methods. 

In our study, we utilize this transparency to turn off the gap plasmon occupation. As effective plasmo-mechanical coupling is suppressed, the noise-squeezing of the mechanical oscillator is also turned off.  When the excitation frequency $\omega$ matches the transparency window, i.e., $\omega = \omega_{\rm f}\sim \Omega_{\rm \scriptscriptstyle QE}$~\cite{PSwf}, the gap intensity turns off. Thus, plasmomechanical interaction is closed and no mechanical displacement appears. However, when $\Omega_{\rm \scriptscriptstyle QE}$ is slightly shifted via an applied voltage~\cite{larocque2024tunable,shibata2013large,miller1985electric}, mechanical displacement~(Fig.~4) and squeezing~(Fig.~5) can be tuned continuously at a desired value.

It is worth nothing that experiments clearly demonstrate that plasmonic structures can handle quantum features, such as entanglement and noise-squeezing, much longer times compared to their typical damping times~\cite{fasel2006quantum,huck2009demonstration,varro2011hanbury,di2012quantum,fasel2005energy,tame2013quantum}. The latter damping is defined for flowing out of the field from the MNS while former governs the degrading of the quantum noise. In other words, degrading time for the squeezed-noise~(solely determines the quantum optics features~\cite{simon1994quantum}) are much longer.

The paper is organized as follows. In Sec. II, we briefly mention about Fano resonances: how and why they appear in plasmonic spectra. In Sec. III, we present our finite difference time domain~(FDTD) simulations. We show that the ordinarily very strong gap plasmon field is suppressed at $\omega=\omega_{\rm f}\sim \Omega_{\rm \scriptscriptstyle QE}$ by the presence of a defect-center layer. The intensity at the gap can be continuously tuned by slightly changing $\Omega_{\rm \scriptscriptstyle QE}$. In Sec. IV, we study the hamiltonian governing the plasmo-mechanical interaction in the presence of the Fano resonance. We calculate the noise-squeezing in the mechanical motion using Langevin equations for the noise operators, e.g., $\delta\hat{a}$~\cite{genes2008robust,vitali2007optomechanical}. We show that the noise-squeezing can also be tuned continuously.



\section{Fano resonances} \label{sec:FanoResonance}

Coupling of a quantum object~(QE) to a plasmon excitation introduces two paths for the absorption of the incident field. The two paths destructively interfere with each other and the absorption of incident field is canceled~\cite{alzar2001classical,tacsgin2013metal}. Thus, a transparency in the plasmonic spectrum appears. This phenomenon can be demonstrated on a single equation~\cite{tacsgin2018fano}
\begin{equation}
	\alpha_p=\frac{\varepsilon_p }{[i(\Omega_p-\omega)+\gamma_p]   - \frac{y|f|^2}{ i(\Omega_{\rm \scriptscriptstyle QE}-\omega) + \gamma_{\rm \scriptscriptstyle QE} } }.
	\label{FR}
\end{equation} 
Here, $\alpha_p$ is the plasmon amplitude. The plasmon mode, of resonance $\Omega_p$~($\lambda_p=c/\Omega_p$) and damping rate $\gamma_p$, is pumped by an incident field of frequency $\omega$ and amplitude $\varepsilon_p$. The QE has a level-spacing of $\Omega_{\rm \scriptscriptstyle QE}$ and a decay rate $\gamma_{\rm \scriptscriptstyle QE}$. The coupling between the QE and the plasmon excitation, $f$, is highly enhanced due to the field localization at the hotspot. $y$ is the population inversion.
$\gamma_{\rm \scriptscriptstyle QE} \simeq 10^{10}$ Hz is 4 orders of smaller than the plasmonic one $\gamma_p \sim 10^{14}$ Hz. $\omega$ and $\Omega_p$ are at optical frequencies $\sim 10^{15}$ Hz. Thus, in frequencies scaled with optical ones, $\omega\sim 1$, $\Omega_p\sim 1$, $\Omega_{\rm \scriptscriptstyle QE}~\sim 1$,  $\gamma_p\sim 0.1$, $\gamma_{\rm \scriptscriptstyle QE}\sim 10^{-5}$ and $f\simeq 0.1$~\cite{singh2016enhancement,wu2010quantum,leng2018strong}.  

When the incident field $\omega$ is in resonance with the QE, i.e., $\omega\simeq \Omega_{\rm \scriptscriptstyle QE}$, the $\gamma_{\rm \scriptscriptstyle QE}\sim 10^{-5}$ leaves alone in the second term of the denominator. It comes out as a factor of $10^5$ that multiplies the $y|f|^2$ term. Thus, the second term of the denominator becomes exceeding large compared to the first term. This suppresses the plasmon excitation amplitude $\alpha$ which creates a transparency at $\omega=\Omega_{\rm \scriptscriptstyle QE}$. 

The transparency takes place in a very narrow band for small $\gamma_{\rm \scriptscriptstyle QE}$. We propose to tune $\Omega_{\rm \scriptscriptstyle QE}$ via an applied voltage~\cite{larocque2024tunable,miller1985electric,shibata2013large} which changes the gap intensity substantially. The experiments shows that 20 meV tuning in $\Omega_{\rm \scriptscriptstyle QE}$ can easily be achieved only with a 1 V of applied potential difference~\cite{larocque2024tunable,shibata2013large}.

We note that Eq.~(\ref{FR}) is derived from an analytical model which cannot account the retardation effects. That is, Eq.~(\ref{FR}) is obtained
as if all interactions with the normally extended plasmon field takes place at a single point. Retardation effects shift the spectral position of the FR, $\omega_{\rm f}$, more with respect to $\Omega_{\rm \scriptscriptstyle QE}$ as LSP gap size becomes larger.
 So in the next section, we present a realistic demonstration of the phenomenon for a spatially extended plasmon profile. In Sec.~IV, we calculate the entanglement/squeezing in a plasmomechanical system controlled by a FR.


\section{FDTD Simulations} \label{sec:FDTD}
 
  In this section, we provide  exact solutions of the 3D Maxwell equations for a sample plasmo-mechanical system, i.e., similar to the one experimentally studied in Ref.~\cite{roxworthy2018electrically}. A cuboid gold nanoparticle of dimensions 20$\times$20$\times$40 nm is embedded into the lower face of a cantilever made of silicon nitride. There lies a gold pad 20 nm below the cantilever~(also the cuboid). The gap between the two gold surfaces ---the cuboid and the gold pad--- supports a gap localized surface plasmon~(LSP) resonance~(LSPR). The LSP is excited by a z-polarized light. The gap plasmons interact with the cantilever~(mechanical system) that supports mechanical oscillations at about $\omega_m=$10 MHz. Zero point fluctuations~(ZPF) of such a cantilever is reported to be about $x_{\rm \scriptscriptstyle ZPF}\simeq $30 fm in Ref.~\cite{roxworthy2018electrically}. Thus, control of the gap intensity~(See Fig.~2b) should be expected to tune the mechanical displacement in units of 30 fm~\cite{PS2}.

 First, we study the spectrum of the LSPRs. We carry out FDTD simulations in Lumerical using experimental dielectric functions for the silicon nitride and gold.   In Fig.~2a we observe that the system supports two LSPRs centered at 533 nm and $\lambda_p=$610 nm. We, as an instance, choose to study at the resonance $\lambda_p=$610 nm. In Fig. 2b, we plot the gap intensity profile, for instance, at $\lambda=$598 nm. Here we present the intensity profile at $\lambda=$598 nm~(i.e., not at the peak) merely because Fano transparency appears at $\lambda_{\rm f}=$598 nm in Fig.~3a when we place the defect-center layer. We aim a comparison.

 \begin{figure}
 	\centering
 	\includegraphics[width=0.98\linewidth]{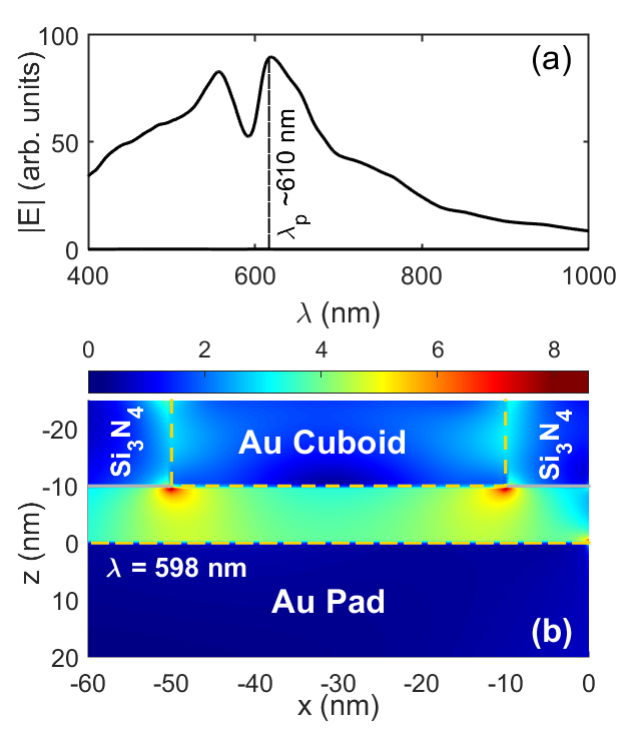}
 	\newline
 	\caption{(a) The plasmonic gap in Fig.~1 supports two LSP resonances at 533 nm and $\lambda_p=$610 nm. (b) We plot the gap intensity for $\lambda=$598 nm~(i.e., not at the plasmonic peak) as we will compare it with the case when QE layer is present. In Fig.~3, Fano transparency appears at $\omega=\omega_f=$598 nm.}
 	\label{fig2}
 \end{figure} 
 
 After determining the LSPRs and the gap intensity profile, we aim to provide a simple demonstration of the Fano resonance~(transparency) in such a system. We place a 3-nm-thick layer of QE standing for the densely-implanted defect-centers~\cite{murata2011high,rotem2007enhanced,ngandeu2024hot}. The layer is positioned on the gold pad and simulated by a Lorentzian dielectric function~\cite{wu2010quantum,shah2013ultrafast,leng2018strong} of resonance $\Omega_{\rm \scriptscriptstyle QE}=610$ nm, linewidth $\gamma_{\rm \scriptscriptstyle QE}=10^{11}$ Hz  and oscillator strength $f_{\rm  osc}=0.1$~\cite{wu2010quantum,leng2018strong}. 
 
In Fig.~3a we plot the spectrum obtained from a point monitor positioned 1 nm below the cuboid. We choose an increased number of frequency points in order to see if there appears a narrow transparency window. We observe that a transparency window appears at about $\lambda_{\rm f}=598$ nm which is not at the $\Omega_{\rm \scriptscriptstyle QE}=610$ nm. This is something to be expected as Eq.~(1) is valid when retardation effects are ignored. This negligence certainly does not hold for a large plasmonic gap of 20 nm width~\cite{PSwf}.

In Fig.~3b, we plot the gap intensity profile at $\lambda=\lambda_{\rm f}=$598 nm. One can observe that gap intensity, so the plasmomechanical coupling, is suppressed substantially  compared to Fig.~2b.

\begin{figure}
	\centering
	\includegraphics[width=0.98\linewidth]{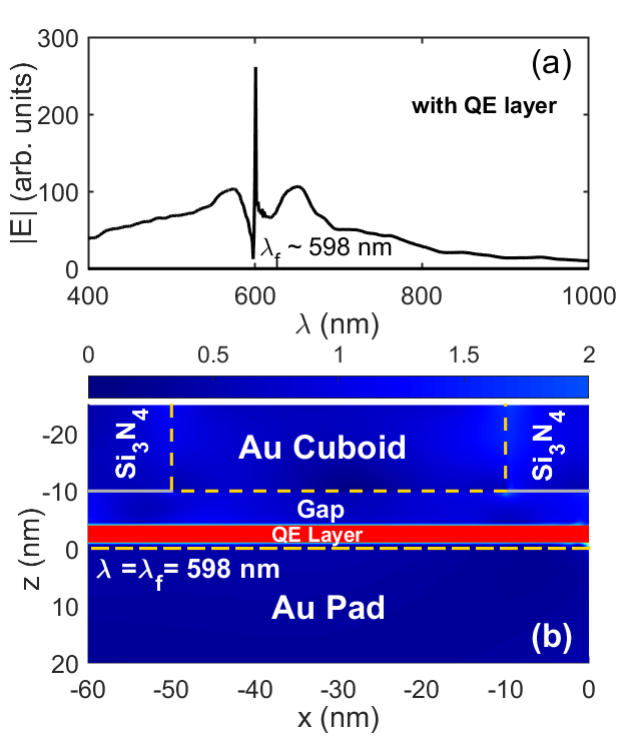}
	\newline
	\caption{(a) The defect layer of resonance $\Omega_{\rm \scriptscriptstyle QE}=$610 nm introduces a Fano resonance~(transparency) at $\omega=\omega_{\rm f}=$598 nm. The shift from $\Omega_{\rm \scriptscriptstyle QE}$ is due to retardation effect~\cite{PSwf}. (b) The FR turns off the gap intensity so the effective plasmomechanical coupling. Thus, by changing the defect resonance $\Omega_{\rm \scriptscriptstyle QE}$, via an applied voltage, one can tune the gap intensity at a fixed driving frequency, e.g., 598 nm. See Fig.~4. The peak at 602 nm is Fano enhancement effect~\cite{tacsgin2018fano,singh2016enhancement}.}
	\label{fig3}
\end{figure}

Next, we slightly shift $\Omega_{\rm \scriptscriptstyle QE}$ step by step assuming that a varying voltage is applied on the defect layer. In Fig.~4 we show that mean gap intensity can be tuned between 120$I_0$ and 30.000$I_0$.  Thus, our FDTD simulations clearly demonstrate that gap intensity, so the effective plasmomechanical coupling and mechanical displacement~(in units of $x_{\rm \scriptscriptstyle ZPF}\simeq 30$ fm)~\cite{genes2008robust,vitali2007optomechanical} can be continuously tuned by an applied voltage of only 0.8 V ---the potential difference needed to tune $\Omega_{\rm \scriptscriptstyle QE}$ from 2.035 eV to 2.042 eV~\cite{larocque2024tunable,shibata2013large,miller1985electric} in Fig.~4.

\begin{figure}
	\centering
	\includegraphics[width=0.98\linewidth]{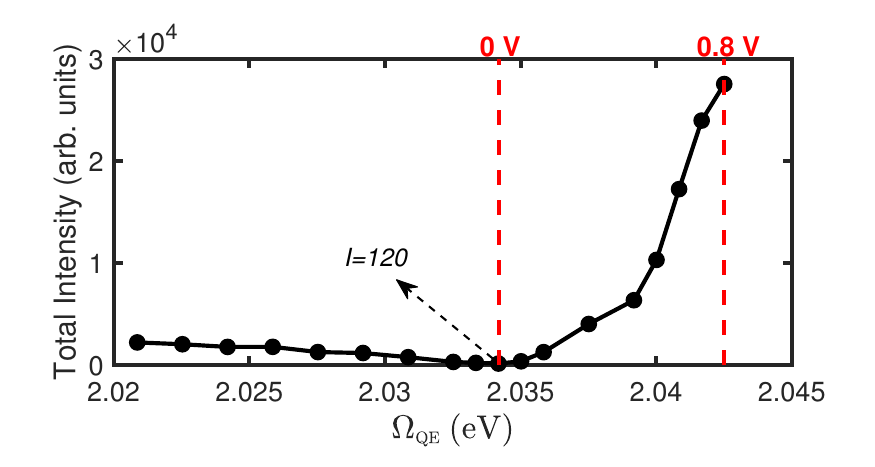}
	\newline
	\caption{ Continuous electrical-tuning of the gap intensity by shifting the Fano resonance via applied voltage~\cite{larocque2024tunable,shibata2013large}. One volt potential difference is more than enough to tune the gap intensity between 120$I_0$--30.000$I_0$~(or 200 times). Plasmomechanical coupling, cantilever displacement and entanglement is tuned in the same manner~\cite{genes2008robust,vitali2007optomechanical}. }
	\label{fig4}
\end{figure}


\section{Tuning noise-squeezing and entanglement} \label{sec:squeezing}

Electrical tuning is also possible for quantum optical features of the Fano-controlled plasmomechanical system. Here, we show that displacement $x_m$, Fig.~5a, and the quadrature-squeezing of the mechanical motion, Fig.~5b, can also be continuously tuned via the applied voltage. The mechanical squeezing can either be used to reduce the uncertainty in the  cantilever motion below the standard quantum limit, i.e., $x_{\rm \scriptscriptstyle ZPF}\simeq 30$ fm in Ref.~\cite{roxworthy2018electrically}, or  to generate entanglement between different oscillator(s) modes~\cite{ge2015conservation,von2024engineering}.

%
%

Dynamics of the system can be summarized as follows. Incident light of frequency $\omega$ drives the gap LSPs. Incident field~(intensity $I_0$) is localized at the hotspot where the local plasmon intensity becomes $\sim 10^4\times I_0$ times the incident one. See Fig. 2a. The  defect-center layer is placed into the gap. It lies on the gold pad. Thus, QE interacts with the plasmon mode strongly. Its interaction with the incident field becomes very weak compared to the plasmons. (Note the 4 orders-of-magnitude ratio between the two intensities.) The gap plasmon intensity also applies a radiation pressure that induces a coupling between the mechanical vibrations~$\hat{x}_m=(\hat{a}_m^\dagger + \hat{a}_m)/\sqrt{2}$ and the plasmon mode $\hat{a}$. 

Total hamiltonian can be written as
\begin{eqnarray}
{\cal \hat{H}} = \hbar\omega_m \hat{a}_m^\dagger\hat{a}_m + \hbar\Omega \hat{a}^\dagger \hat{a} + \hbar \varepsilon_p (\hat{a}^\dagger e^{-i\omega t} + \hat{a} e^{i\omega t} )
\\ \nonumber
+  \hbar \omega_{eg} |e \rangle\langle e|   +  \hbar g \hat{a}^\dagger\hat{a} \hat{x}_m + \hbar f (\hat{a} |e \rangle\langle g| + \hat{a}^\dagger |g \rangle\langle e| ),
\end{eqnarray}
where the terms are, respectively, energy of the mechanical oscillations, plasmons, the coupling to the incident field, excitation energy of the QE, radiation pressure interaction and QE-plasmon coupling that induces the Fano transparency.

We calculate the entanglement features using the standard methods~\cite{genes2008robust,vitali2007optomechanical} that can be found in the literature. See the supplementary material~(SM)~\cite{supp}. In Fig.~5a, we present the expectation value of the cantilever displacement in units of $x_{\rm \scriptscriptstyle ZPF}\simeq$30 fm given in Ref.~\cite{roxworthy2018electrically}. Fig.~5b plots the nonclassicality~(e.g., squeezing) in units of its entanglement potential~\cite{asboth2005computable}. This quantifies how much entanglement is generated when a single squeezed mode is splitted into two in a beam-splitter. The applied voltage tunes the gap intensity, so the effective coupling $G= \alpha_p \times g$~\cite{genes2008robust,vitali2007optomechanical}. Smaller $G$ results in smaller displacement and entanglement. We use the parameters of Ref.~\cite{roxworthy2018electrically}.

\begin{figure}
	\centering
	\includegraphics[width=0.9\linewidth]{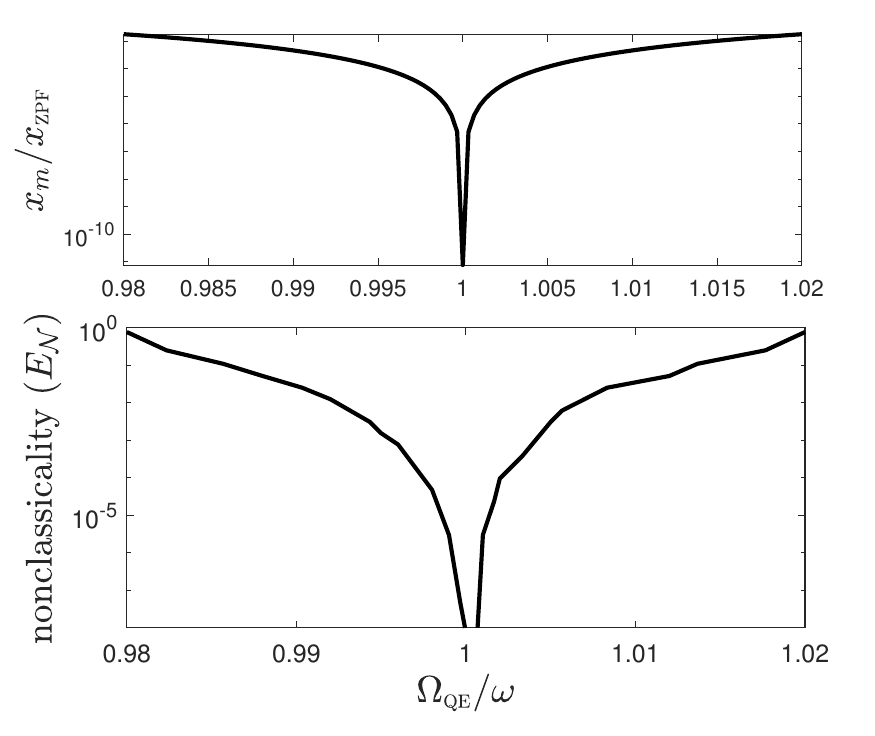}
	\newline
	\caption{(a) Displacement and (b) quantumness of the oscillator motion can also be electrically tuned in a continuous manner. Quantumness~(e.g. squeezing) of the mechanical motion is quantified in terms of log-neg~\cite{plenio2005logarithmic,vidal2002computable}, entanglement such a single-mode generates when it is divided into two at a beam-splitter~\cite{tasgin2020measuring,asboth2005computable}. Displacement is scaled with $x_{\rm \scriptscriptstyle ZPF}=$30 fm given in Ref.~\cite{roxworthy2018electrically}. }
	\label{fig5}
\end{figure}


\section{Summary and Outlooks} \label{sec:summary}

In summary, we provide a device where we can control the plasmo-mechanical coupling by a factor of 200. We do not tilt the cantilever equilibrium position. Instead, we gradually turn on and off the gap plasmon intensity via an applied voltage. This is not only faster~(ps) than tuning the driving optical field intensity, but also provides enhanced integrability.
 (In order to appreciate, please consider a number of these devices integrated at different places to a circuit. One would need to tune each driving intensity separately that would necessitate much larger spaces. In our method, however, a common drive, i.e., a waveguide, space for light would be sufficient.) 
 
We utilize the Fano transparency induced by a quantum object~(layer of defect-centers) in order to tune the gap intensity. This way, we can achieve very high precision cantilever displacements (units of 30 fermimeters)  without bending the cantilever that limits the precision with nm. Moreover, we can controllably tune~(reduce) the noise in the mechanical displacement $x_m$. On top of this, the typical response time for such a tuning in our system is in picoseconds, thus limited only by the speed of the device tuning the applied voltage. Such a fast and  precision controlled displacement cannot be achieved with other configuration to our best knowledge. Naturally, one needs sideband cooling~\cite{bhattacharya2007trapping} at room temperature or needs to work at cryogenic temperatures in order to achieve such precisions. 


We should renote that such precise displacement tunings are already available in optomechanics at room temperature via sideband cooling technique~\cite{bhattacharya2007trapping}. However, one controls the pump intensity. 

%

\bibliography{bibliography}	

\begin{thebibliography}{50}
\providecommand{\natexlab}[1]{#1}
\providecommand{\url}[1]{\texttt{#1}}
\expandafter\ifx\csname urlstyle\endcsname\relax
  \providecommand{\doi}[1]{doi: #1}\else
  \providecommand{\doi}{doi: \begingroup \urlstyle{rm}\Url}\fi

\bibitem[Bhattacharya and Meystre(2007)]{bhattacharya2007trapping}
M~Bhattacharya and P~Meystre.
\newblock Trapping and cooling a mirror to its quantum mechanical ground state.
\newblock \emph{Physical Review Letters}, 99\penalty0 (7):\penalty0 073601,
  2007.

\bibitem[Zhang et~al.(2003)Zhang, Peng, and Braunstein]{zhang2003quantum}
Jing Zhang, Kunchi Peng, and Samuel~L Braunstein.
\newblock Quantum-state transfer from light to macroscopic oscillators.
\newblock \emph{Physical Review A}, 68\penalty0 (1):\penalty0 013808, 2003.

\bibitem[Meystre(2013)]{meystre2013short}
Pierre Meystre.
\newblock A short walk through quantum optomechanics.
\newblock \emph{Annalen der Physik}, 525\penalty0 (3):\penalty0 215--233, 2013.

\bibitem[Jia et~al.(2024)Jia, Xu, Kuns, Nakano, Barsotti, Evans, Mavalvala,
  Collaboration†, Abbott, Abouelfettouh, et~al.]{jia2024squeezing}
Wenxuan Jia, Victoria Xu, Kevin Kuns, Masayuki Nakano, Lisa Barsotti, Matthew
  Evans, Nergis Mavalvala, LIGO~Scientific Collaboration†, R~Abbott,
  I~Abouelfettouh, et~al.
\newblock Squeezing the quantum noise of a gravitational-wave detector below
  the standard quantum limit.
\newblock \emph{Science}, 385\penalty0 (6715):\penalty0 1318--1321, 2024.

\bibitem[Thijssen et~al.(2013)Thijssen, Verhagen, Kippenberg, and
  Polman]{thijssen2013plasmon}
Rutger Thijssen, Ewold Verhagen, Tobias~J Kippenberg, and Albert Polman.
\newblock Plasmon nanomechanical coupling for nanoscale transduction.
\newblock \emph{Nano Letters}, 13\penalty0 (7):\penalty0 3293--3297, 2013.

\bibitem[Roxworthy and Aksyuk(2016)]{roxworthy2016nanomechanical}
Brian~J Roxworthy and Vladimir~A Aksyuk.
\newblock Nanomechanical motion transduction with a scalable localized gap
  plasmon architecture.
\newblock \emph{Nature Communications}, 7\penalty0 (1):\penalty0 13746, 2016.

\bibitem[Ou et~al.(2016)Ou, Plum, Zhang, and Zheludev]{ou2016giant}
Jun-Yu Ou, Eric Plum, Jianfa Zhang, and Nikolay~I Zheludev.
\newblock Giant nonlinearity of an optically reconfigurable plasmonic
  metamaterial.
\newblock \emph{Advanced Materials}, 28\penalty0 (4):\penalty0 729--733, 2016.

\bibitem[Zhu et~al.(2016)Zhu, Yi, and Cubukcu]{zhu2016plasmonic}
Hai Zhu, Fei Yi, and Ertugrul Cubukcu.
\newblock Plasmonic metamaterial absorber for broadband manipulation of
  mechanical resonances.
\newblock \emph{Nature Photonics}, 10\penalty0 (11):\penalty0 709--714, 2016.

\bibitem[Dong et~al.(2016)Dong, Chen, Zhou, Wang, Zhang, and
  Sun]{dong2016gigahertz}
Biqin Dong, Xiangfan Chen, Fan Zhou, Chen Wang, Hao~F Zhang, and Cheng Sun.
\newblock Gigahertz all-optical modulation using reconfigurable nanophotonic
  metamolecules.
\newblock \emph{Nano letters}, 16\penalty0 (12):\penalty0 7690--7695, 2016.

\bibitem[Roxworthy and Aksyuk(2018)]{roxworthy2018electrically}
Brian~J Roxworthy and Vladimir~A Aksyuk.
\newblock Electrically tunable plasmomechanical oscillators for localized
  modulation, transduction, and amplification.
\newblock \emph{Optica}, 5\penalty0 (1):\penalty0 71--79, 2018.

\bibitem[Haffner et~al.(2019)Haffner, Joerg, Doderer, Mayor, Chelladurai,
  Fedoryshyn, Roman, Mazur, Burla, Lezec, et~al.]{haffner2019nano}
Christian Haffner, Andreas Joerg, Michael Doderer, Felix Mayor, Daniel
  Chelladurai, Yuriy Fedoryshyn, Cosmin~Ioan Roman, Mikael Mazur, Maurizio
  Burla, Henri~J Lezec, et~al.
\newblock Nano--opto-electro-mechanical switches operated at cmos-level
  voltages.
\newblock \emph{Science}, 366\penalty0 (6467):\penalty0 860--864, 2019.

\bibitem[Xu et~al.(2021)Xu, Cheng, Tang, Lv, Li, Guo, Wang, Song, Zhou, and
  Deng]{xu2021recent}
Nan Xu, Ze-Di Cheng, Jin-Dao Tang, Xiao-Min Lv, Tong Li, Meng-Lin Guo, You
  Wang, Hai-Zhi Song, Qiang Zhou, and Guang-Wei Deng.
\newblock Recent advances in nano-opto-electro-mechanical systems.
\newblock \emph{Nanophotonics}, 10\penalty0 (9):\penalty0 2265--2281, 2021.

\bibitem[Koya et~al.(2021)Koya, Cunha, Guerrero-Becerra, Garoli, Wang,
  Juodkazis, and Proietti~Zaccaria]{koya2021plasmomechanical}
Alemayehu~Nana Koya, Joao Cunha, Karina~Andrea Guerrero-Becerra, Denis Garoli,
  Tao Wang, Saulius Juodkazis, and Remo Proietti~Zaccaria.
\newblock Plasmomechanical systems: principles and applications.
\newblock \emph{Advanced Functional Materials}, 31\penalty0 (41):\penalty0
  2103706, 2021.

\bibitem[Genes et~al.(2008)Genes, Mari, Tombesi, and Vitali]{genes2008robust}
Claudiu Genes, Andrea Mari, Paolo Tombesi, and David Vitali.
\newblock Robust entanglement of a micromechanical resonator with output
  optical fields.
\newblock \emph{Physical Review A—Atomic, Molecular, and Optical Physics},
  78\penalty0 (3):\penalty0 032316, 2008.

\bibitem[Vitali et~al.(2007)Vitali, Gigan, Ferreira, B{\"o}hm, Tombesi,
  Guerreiro, Vedral, Zeilinger, and Aspelmeyer]{vitali2007optomechanical}
David Vitali, Sylvain Gigan, Anderson Ferreira, HR~B{\"o}hm, Paolo Tombesi,
  Ariel Guerreiro, Vlatko Vedral, <?~format?>~A Zeilinger, and Markus
  Aspelmeyer.
\newblock Optomechanical entanglement between a movable mirror and a cavity
  field.
\newblock \emph{Physical review letters}, 98\penalty0 (3):\penalty0 030405,
  2007.

\bibitem[Leng et~al.(2018)Leng, Szychowski, Daniel, and Pelton]{leng2018strong}
Haixu Leng, Brian Szychowski, Marie-Christine Daniel, and Matthew Pelton.
\newblock Strong coupling and induced transparency at room temperature with
  single quantum dots and gap plasmons.
\newblock \emph{Nature Communications}, 9\penalty0 (1):\penalty0 4012, 2018.

\bibitem[Wu et~al.(2010)Wu, Gray, and Pelton]{wu2010quantum}
Xiaohua Wu, Stephen~K Gray, and Matthew Pelton.
\newblock Quantum-dot-induced transparency in a nanoscale plasmonic resonator.
\newblock \emph{Optics Express}, 18\penalty0 (23):\penalty0 23633--23645, 2010.

\bibitem[Shah et~al.(2013)Shah, Scherer, Pelton, and Gray]{shah2013ultrafast}
Raman~A Shah, Norbert~F Scherer, Matthew Pelton, and Stephen~K Gray.
\newblock Ultrafast reversal of a fano resonance in a plasmon-exciton system.
\newblock \emph{Physical Review B—Condensed Matter and Materials Physics},
  88\penalty0 (7):\penalty0 075411, 2013.

\bibitem[Tasgin et~al.(2018)Tasgin, Bek, and Postac{\i}]{tacsgin2018fano}
Mehmet~Emre Tasgin, Alpan Bek, and Selen Postac{\i}.
\newblock Fano resonances in the linear and nonlinear plasmonic response.
\newblock \emph{Fano Resonances in Optics and Microwaves: Physics and
  Applications}, 219, 2018.
\newblock URL
  \url{https://link.springer.com/chapter/10.1007/978-3-319-99731-5_1}.

\bibitem[Limonov et~al.(2017)Limonov, Rybin, Poddubny, and
  Kivshar]{limonov2017fano}
Mikhail~F Limonov, Mikhail~V Rybin, Alexander~N Poddubny, and Yuri~S Kivshar.
\newblock Fano resonances in photonics.
\newblock \emph{Nature Photonics}, 11\penalty0 (9):\penalty0 543--554, 2017.

\bibitem[PS_()]{PS_sideband}
Sideband cooling is not demonstrated for plasmomechanics yet to our knowledge.

\bibitem[Ge et~al.(2015)Ge, Tasgin, and Zubairy]{ge2015conservation}
Wenchao Ge, Mehmet~Emre Tasgin, and M~Suhail Zubairy.
\newblock Conservation relation of nonclassicality and entanglement for
  gaussian states in a beam splitter.
\newblock \emph{Physical Review A}, 92\penalty0 (5):\penalty0 052328, 2015.

\bibitem[von L{\"u}pke et~al.(2024)von L{\"u}pke, Rodrigues, Yang, Fadel, and
  Chu]{von2024engineering}
Uwe von L{\"u}pke, Ines~C Rodrigues, Yu~Yang, Matteo Fadel, and Yiwen Chu.
\newblock Engineering multimode interactions in circuit quantum
  acoustodynamics.
\newblock \emph{Nature Physics}, 20\penalty0 (4):\penalty0 564--570, 2024.

\bibitem[Murata et~al.(2011)Murata, Yasutake, Nittoh, Fukatsu, and
  Miki]{murata2011high}
Koichi Murata, Yuhsuke Yasutake, Koh-ichi Nittoh, Susumu Fukatsu, and Kazushi
  Miki.
\newblock High-density g-centers, light-emitting point defects in silicon
  crystal.
\newblock \emph{AIP Advances}, 1\penalty0 (3), 2011.

\bibitem[Rotem et~al.(2007)Rotem, Shainline, and Xu]{rotem2007enhanced}
Efraim Rotem, Jeffrey~M Shainline, and Jimmy~M Xu.
\newblock Enhanced photoluminescence from nanopatterned carbon-rich silicon
  grown by solid-phase epitaxy.
\newblock \emph{Applied Physics Letters}, 91\penalty0 (5), 2007.

\bibitem[Ngandeu~Ngambou et~al.(2024)Ngandeu~Ngambou, Perrin, Balasa, Tiranov,
  Brinza, Bénédic, Renaud, Reveillard, Silvent, Goldner, Achard, and
  Tallaire]{ngandeu2024hot}
Midrel~Wilfried Ngandeu~Ngambou, Pauline Perrin, Ionut Balasa, Alexey Tiranov,
  Ovidiu Brinza, Fabien Bénédic, Justine Renaud, Morgan Reveillard, Jérémie
  Silvent, Philippe Goldner, Jocelyn Achard, and Alexandre Tallaire.
\newblock Hot ion implantation to create dense {NV} center ensembles in
  diamond.
\newblock \emph{Applied Physics Letters}, 124\penalty0 (13):\penalty0 134002,
  03 2024.
\newblock ISSN 0003-6951.
\newblock \doi{10.1063/5.0196719}.

\bibitem[PSw()]{PSwf}
Please note that $\omega_{\rm f}=\Omega_{\rm \scriptscriptstyle QE}$ when all
  interactions are assumed to take place at a single point, i.e., as assumed in
  the derivation of Eq.~(1). When spatial profile of the LSP is large, e.g., in
  case of a 20 nm width gap, retardation effects make the Fano resonance shift
  substantially as we observe in Fig.~3a. Larger the LSP gap size, $\omega_{\rm
  f}$ shifts more with respect to $\Omega_{\rm \scriptscriptstyle QE}$.

\bibitem[Larocque et~al.(2024)Larocque, Buyukkaya, Errando-Herranz, Papon,
  Harper, Tao, Carolan, Lee, Richardson, Leake, et~al.]{larocque2024tunable}
Hugo Larocque, Mustafa~Atabey Buyukkaya, Carlos Errando-Herranz, Camille Papon,
  Samuel Harper, Max Tao, Jacques Carolan, Chang-Min Lee, Christopher~JK
  Richardson, Gerald~L Leake, et~al.
\newblock Tunable quantum emitters on large-scale foundry silicon photonics.
\newblock \emph{Nature Communications}, 15\penalty0 (1):\penalty0 5781, 2024.

\bibitem[Shibata et~al.(2013)Shibata, Yuan, Iwasa, and
  Hirakawa]{shibata2013large}
Kenji Shibata, Hongtao Yuan, Yoshihiro Iwasa, and Kazuhiko Hirakawa.
\newblock Large modulation of zero-dimensional electronic states in quantum
  dots by electric-double-layer gating.
\newblock \emph{Nature Communications}, 4\penalty0 (1):\penalty0 2664, 2013.

\bibitem[Miller et~al.(1985)Miller, Chemla, Damen, Gossard, Wiegmann, Wood, and
  Burrus]{miller1985electric}
David~AB Miller, DS~Chemla, TC~Damen, AC~Gossard, W~Wiegmann, TH~Wood, and
  CA~Burrus.
\newblock Electric field dependence of optical absorption near the band gap of
  quantum-well structures.
\newblock \emph{Physical Review B}, 32\penalty0 (2):\penalty0 1043, 1985.

\bibitem[Fleischhauer et~al.(2005)Fleischhauer, Imamoglu, and
  Marangos]{fleischhauer2005electromagnetically}
Michael Fleischhauer, Atac Imamoglu, and Jonathan~P Marangos.
\newblock Electromagnetically induced transparency: Optics in coherent media.
\newblock \emph{Reviews of Modern Physics}, 77\penalty0 (2):\penalty0 633--673,
  2005.

\bibitem[Ta{\c{s}}g{\i}n(2013)]{tacsgin2013metal}
Mehmet~Emre Ta{\c{s}}g{\i}n.
\newblock Metal nanoparticle plasmons operating within a quantum lifetime.
\newblock \emph{Nanoscale}, 5\penalty0 (18):\penalty0 8616--8624, 2013.

\bibitem[PSD()]{PSDarkmodes}
Fano resonances also appear when a dark plasmon mode, narrower compared to a
  bright mode, is coupled to a bright plasmon mode. This phenomenon, usually
  studied in the literature, is not to be confused with the FRs that appear due
  to QEs. The former ones are not so useful as the resonances of the dark
  plasmon modes cannot actively be tuned, e.g, via applied voltage. Moreover,
  FRs with dark modes are much broader compared to the ones with QEs.

\bibitem[G{\"u}nay et~al.(2020)G{\"u}nay, Chuang, and
  Tasgin]{gunay2020continuously}
Mehmet G{\"u}nay, You-Lin Chuang, and Mehmet~Emre Tasgin.
\newblock Continuously-tunable cherenkov-radiation-based detectors via plasmon
  index control.
\newblock \emph{Nanophotonics}, 9\penalty0 (6):\penalty0 1479--1489, 2020.

\bibitem[Polat et~al.(2024)Polat, Artvin, {\c{S}}aki, Bek, and
  Sahin]{polat2024continuous}
Emre~Ozan Polat, Zafer Artvin, Yusuf {\c{S}}aki, Alpan Bek, and Ramazan Sahin.
\newblock Continuous and reversible electrical tuning of fluorescent decay rate
  via fano resonance.
\newblock \emph{arXiv preprint arXiv:2412.20199}, 2024.

\bibitem[Fasel et~al.(2006)Fasel, Halder, Gisin, and Zbinden]{fasel2006quantum}
Sylvain Fasel, Matth{\"a}us Halder, Nicolas Gisin, and Hugo Zbinden.
\newblock Quantum superposition and entanglement of mesoscopic plasmons.
\newblock \emph{New Journal of Physics}, 8\penalty0 (1):\penalty0 13, 2006.

\bibitem[Huck et~al.(2009)Huck, Smolka, Lodahl, S{\o}rensen, Boltasseva,
  Janousek, and Andersen]{huck2009demonstration}
Alexander Huck, Stephan Smolka, Peter Lodahl, Anders~S S{\o}rensen, Alexandra
  Boltasseva, <?~format?>~Jiri Janousek, and Ulrik~L Andersen.
\newblock Demonstration of quadrature-squeezed surface plasmons in a gold
  waveguide.
\newblock \emph{Physical Review Letters}, 102\penalty0 (24):\penalty0 246802,
  2009.

\bibitem[Varr{\'o} et~al.(2011)Varr{\'o}, Kro{\'o}, Oszetzky, Nagy, and
  Czitrovszky]{varro2011hanbury}
S{\'a}ndor Varr{\'o}, Norbert Kro{\'o}, D{\'a}niel Oszetzky, Attila Nagy, and
  Alad{\'a}r Czitrovszky.
\newblock Hanbury brown--twiss type correlations with surface plasmon light.
\newblock \emph{Journal of Modern Optics}, 58\penalty0 (21):\penalty0
  2049--2057, 2011.

\bibitem[Di~Martino et~al.(2012)Di~Martino, Sonnefraud, K{\'e}na-Cohen, Tame,
  Ozdemir, Kim, and Maier]{di2012quantum}
Giuliana Di~Martino, Yannick Sonnefraud, St{\'e}phane K{\'e}na-Cohen, Mark
  Tame, Sahin~K Ozdemir, MS~Kim, and Stefan~A Maier.
\newblock Quantum statistics of surface plasmon polaritons in metallic stripe
  waveguides.
\newblock \emph{Nano Letters}, 12\penalty0 (5):\penalty0 2504--2508, 2012.

\bibitem[Fasel et~al.(2005)Fasel, Robin, Moreno, Erni, Gisin, and
  Zbinden]{fasel2005energy}
Sylvain Fasel, Franck Robin, Esteban Moreno, Daniel Erni, Nicolas Gisin, and
  Hugo Zbinden.
\newblock Energy-time entanglement preservation in plasmon-assisted light
  transmission.
\newblock \emph{Physical Review Letters}, 94\penalty0 (11):\penalty0 110501,
  2005.

\bibitem[Tame et~al.(2013)Tame, McEnery, {\"O}zdemir, Lee, Maier, and
  Kim]{tame2013quantum}
Mark~S Tame, KR~McEnery, {\c{S}}K~{\"O}zdemir, Jinhyoung Lee, Stefan~A Maier,
  and MS~Kim.
\newblock Quantum plasmonics.
\newblock \emph{Nature Physics}, 9\penalty0 (6):\penalty0 329--340, 2013.

\bibitem[Simon et~al.(1994)Simon, Mukunda, and Dutta]{simon1994quantum}
Rajiah Simon, Narasimhaiengar Mukunda, and Biswadeb Dutta.
\newblock Quantum-noise matrix for multimode systems: U (n) invariance,
  squeezing, and normal forms.
\newblock \emph{Physical Review A}, 49\penalty0 (3):\penalty0 1567, 1994.

\bibitem[Garrido~Alzar et~al.(2002)Garrido~Alzar, Martinez, and
  Nussenzveig]{alzar2001classical}
C.~L. Garrido~Alzar, M.~A.~G. Martinez, and P.~Nussenzveig.
\newblock Classical analog of electromagnetically induced transparency.
\newblock \emph{American Journal of Physics}, 70\penalty0 (1):\penalty0 37--41,
  01 2002.
\newblock ISSN 0002-9505.
\newblock \doi{10.1119/1.1412644}.

\bibitem[Singh et~al.(2016)Singh, Abak, and Tasgin]{singh2016enhancement}
Shailendra~K Singh, M~Kurtulus Abak, and Mehmet~Emre Tasgin.
\newblock Enhancement of four-wave mixing via interference of multiple
  plasmonic conversion paths.
\newblock \emph{Physical Review B}, 93\penalty0 (3):\penalty0 035410, 2016.

\bibitem[PS2()]{PS2}
Of course, either sideband cooling or working at cryogenic temperatures is
  necessary for such accuracies.

\bibitem[sup()]{supp}
See Supplemental Material at URL-will-be-inserted-by-publisher.

\bibitem[Asb{\'o}th et~al.(2005)Asb{\'o}th, Calsamiglia, and
  Ritsch]{asboth2005computable}
J{\'a}nos~K Asb{\'o}th, John Calsamiglia, and Helmut Ritsch.
\newblock Computable measure of nonclassicality for light.
\newblock \emph{Physical review letters}, 94\penalty0 (17):\penalty0 173602,
  2005.

\bibitem[Plenio(2005)]{plenio2005logarithmic}
Martin~B Plenio.
\newblock Logarithmic negativity: a full entanglement monotone that is not
  convex.
\newblock \emph{Physical Review Letters}, 95\penalty0 (9):\penalty0 090503,
  2005.

\bibitem[Vidal and Werner(2002)]{vidal2002computable}
Guifr{\'e} Vidal and Reinhard~F Werner.
\newblock Computable measure of entanglement.
\newblock \emph{Physical Review A}, 65\penalty0 (3):\penalty0 032314, 2002.

\bibitem[Tasgin(2020)]{tasgin2020measuring}
Mehmet~Emre Tasgin.
\newblock Measuring nonclassicality of single-mode systems.
\newblock \emph{Journal of Physics B: Atomic, Molecular and Optical Physics},
  53\penalty0 (17):\penalty0 175501, 2020.

\end{thebibliography}
\end{document}